\documentstyle[prb,aps,amsfonts,amstex,12pt]{revtex}

\tighten

\renewcommand{\thanks}[1]{\renewcommand{\thefootnote}{\rm\alph{footnote})}%
   \protect\footnote{#1}}

\makeatletter
\title{Lattice Simulations of the Quantum Microcanonical Ensemble}
\author{
J.~Lukkarinen\thanks{E-mail address: jani.lukkarinen@helsinki.fi}}
\address{ Helsinki Institute of Physics, P.O.Box 9, 
00014 University of Helsinki, Finland} 
\makeatother

\newcommand{\xcite}[1]{%
$\!\!^{\scriptstyle\protect\onlinecite{#1}}$}

\newcommand{\ham}{\widehat{H}}
\newcommand{\tihep}{\widehat{\rho}_\varepsilon}
\newcommand{\half}{ {1\over 2}}
\newcommand{\tr}{{\rm Tr}\,}
\newcommand{\vep}{\varepsilon}
\newcommand{\ci}{{\rm i}}
\newcommand{\re}{{\rm Re\,}} 
\newcommand{\comment}[1]{}

\newtheorem{theorem}{Theorem}[section]
\newtheorem{lemma}[theorem]{Lemma}

\begin{document} 

\preprint{HIP-1997-15/TH}

\maketitle

\begin{abstract}
 We propose a method for the numerical computation of 
microcanonical expectation
values---i.e.\ averages over energy eigenstates with the same
eigenvalue---without any prior knowledge about the spectrum of
the Hamiltonian.  This is accomplished by first defining
a new Gaussian ensemble to approximate the microcanonical
one.  We then develop
a ``lattice theory'' for this Gaussian ensemble
and propose a Monte Carlo integration of the 
expectation values of the lattice theory.
\end{abstract}


\section{Introduction}

The microcanonical ensemble is considered to be 
the fundamental ensemble of statistical physics.  For example, the 
use of the canonical Gibbs ensemble is usually justified by
showing that its expectation values coincide with the microcanonical ones 
in the infinite-volume limit.
However, since an application of 
the microcanonical ensemble requires detailed knowledge about the energy
levels of the system, it is seldom possible to use it in practice.
The canonical ensemble, on the other hand, has a
well-behaved path-integral expression easily extended to
the gauge field theories---therefore it has become the standard
ensemble of quantum statistics.

Nevertheless, there are situations where
the easiest alternative does not work. For instance, direct applications
of the grand canonical ensemble have not been able to reproduce all 
the results of the relativistic ion collision 
experiments.\xcite{redlich:94}
There are two possible explanations for this shortcoming of the canonical
ensemble: either the particle gas created in the collision
does not reach thermal equilibrium before exploding into the final state
hadrons, or the system is too small to be handled by the canonical
methods.

Recent calculations\xcite{braunm:96} using 
the grand canonical ensemble with finite-volume corrections
have, however, succeeded in describing most of the particle abundances in
heavy ion collisions.  This suggests that at least the final state hadron
gas will thermalize, but it also shows that the finite-volume effects are 
prominent in these systems.  

These results point into the direction that the microcanonical
ensemble---which is optimal for describing isolated ergodic systems with
small quantum numbers---should be used for getting quantitative
information about the properties of the quark-gluon plasma possibly
created in the relativistic hadron collisions.  The 
continuum path integral formulation of
the microcanonical ensemble and the corresponding perturbation theory
have already been developed both for
scalar\xcite{chs:93} and gauge field theories,\xcite{chss:94} but since
the phenomena involving hadrons happen in the low energy, strong
coupling regime, non-perturbative methods are also needed.  However, no
general numerical treatment is presently available to compute
microcanonical expectation values without {\em a priori}\/ knowledge
about the energy spectrum even in the quantum mechanical context, let
alone in a quantum field theory.  In this paper, we aim at a development
of such a method in the spirit of lattice field theories for a class of
quantum mechanical Hamiltonians.  The generalization of these results to
the more interesting case of gauge field theories, such as quantum
chromodynamics, is a subject of further work.

The paper is divided into three parts: the first section introduces a
new statistical ensemble, called a Gaussian ensemble, and establishes its
close connection to the microcanonical one.  In the second section
we will derive lattice integral kernels for
a numerical evaluation of the Gaussian expectation values.  Due to the
complicated nature of these kernels, it is not immediately evident how
Monte Carlo methods can be used in this computation---this will be the
subject of the rest of the paper.

\section{The Gaussian ensemble\protect\xcite{HETH:87}}

Consider the operator defined by
\begin{equation} \label{e:tihep} 
\tihep(E) = {1\over \sqrt{2 \pi
\varepsilon^2} } \exp\!\!\left[-\half \left( {\ham -E \over \varepsilon}
\right)^2 \right], 
\end{equation} 
where $\ham$ is the Hamiltonian.  This operator is clearly bounded,
self-adjoint and positive.  Let us assume that the spectrum of the
Hamiltonian is discrete with eigenvalues $E_n$, each of which has a
finite number of eigenvectors $\Omega_{n,k}$, $k=1,\ldots,\kappa_n$,
where $\kappa_n$ is the multiplicity of the eigenvalue $E_n$.  With
rather loose assumptions on the eigenvalues ($\lim_{n\to \infty} n^s /
E_n = 0$ for some $s>0$ will suffice if the multiplicities $\kappa_n$
diverge at most polynomially in $n$) $\tihep$ is trace-class and
therefore defines a sensible statistical ensemble via
\begin{equation} 
\label{e:defev} 
\langle\widehat{A}\rangle \equiv 
{\tr(\tihep \widehat{A}) \over \tr \tihep} 
\mbox{, for any observable }\widehat{A}. 
\end{equation} 
Note that if the Gibbs ensemble is well-defined for temperature
$2\vep^2$, i.e.\ $\tr \exp(-{1\over 2 \vep^2}\ham)<\infty$, then
$E_n-E<1$ only for finitely many $n,k$ and, since for the rest
$(E_n-E)^2 \ge E_n-E$, $\tihep(E)$ is then trace-class.  This
means that the Gaussian ensemble is well-defined for all systems for which
we can use the canonical one.

Intuitively, the above density operator describes a system that
was initially prepared into
a Gaussian energy distribution $E\pm\varepsilon$, but which has then
stabilized under ergodic evolution in isolation.  
This, with the formal relation 
$\lim_{\varepsilon \to 0} \tihep(E) = \delta(\ham - E)$,
will give the physical motivation for using this ensemble to approximate
the microcanonical one.  However, it can also be argued that the Gaussian
ensemble is even better suited for describing real experimental
situations than the microcanonical ensemble, since it
naturally incorporates the energy resolution of the experiment
through the parameter $\vep$---in 
any real experiment the total energy of the system (e.g.\ of a particle)
cannot be known exactly, but only within a certain accuracy.
We will now turn to the mathematical justification of these formal
statements.

Using the earlier notations, we can express the trace in (\ref{e:defev}) as
\begin{equation}
\label{e:tracesum}
\tr (\tihep \widehat{A}) = {1\over \sqrt{2 \pi \varepsilon^2} }
\sum_{n,k} \exp\!\left[ -{1\over 2 \vep^2}
 \left( E_n -E \right)^2 \right] \langle\Omega_{n,k}|\widehat{A}|\Omega_{n,k}\rangle. 
\end{equation}
Since
\begin{equation}
 \lim_{\varepsilon \to 0} 
 \exp\!\left[ -{1\over 2 \vep^2} \left( E_n -E \right)^2 \right] =
 0\mbox{, when } E\ne E_n,
\end{equation}
the partition function clearly satisfies
\begin{equation}
\lim_{\varepsilon \to 0} 
\tr \tihep = \sum_n \kappa_n \delta(E-E_n) = Z_{\rm microcan}.
\end{equation}

In the evaluation of the limit of the expectation values, 
we need the result
\begin{equation}
\label{e:elim}
{ 1 \over \sum\limits_{n',k'} \exp\!\left[ {
     -{(E_{n'} -E)^2 - (E_n -E)^2 \over 2 \varepsilon^2} } \right] } 
 \longrightarrow \left\{\begin{array}{l}
   0\mbox{, if }|E_{n'} -E|< |E_n -E|\mbox{ for some }n' \\
   (\sum_{n'} \kappa_{n'}\delta_{|E_{n'} -E|,|E_n -E|})^{-1}\mbox{, otherwise} 
 \end{array}\right. .
\end{equation}
In other words, this limit is zero if $E_n$ is not the eigenvalue nearest to
$E$, it is  $(\kappa_n+\kappa_{n'})^{-1}$ if both $E_n$ and $E_{n'}$ are nearest
eigenvalues (i.e.\ if $E$ lies exactly in the middle of the segment joining
$E_n$ and $E_{n'}$ and no other eigenvalues are on this segment) and it is 
$\kappa_n^{-1}$ if $E_n$ is a unique nearest eigenvalue.  Let us use
notation $M_n(E)$ for the sum 
$\sum_{n'} \kappa_{n'}\delta_{|E_{n'} -E|,|E_n -E|}$.

Using the formula (\ref{e:tracesum}) for the traces and (\ref{e:elim})
to compute the limit, we arrive at
\begin{equation}
\label{e:meanlim}
\langle\widehat{A}\rangle = \sum_{n,k}
 {\exp\!\left[{ -{1\over 2\vep^2} \left( {E_{n} -E } \right)^2 }\right]
 \over  \tr \tihep \,\sqrt{2 \pi \varepsilon^2} } 
 \langle\Omega_{n,k}| \widehat{A} |\Omega_{n,k}\rangle 
 \longrightarrow {1\over M_{n_0}(E)}
% \sum\limits_{\stackrel{\scriptstyle|E_{n} -E|=}{|E_{n_0} -E|}}
%  \langle\Omega_{n,k}| \widehat{A} |\Omega_{n,k}\rangle,
 \sum_{n,k} \delta_{|E_{n} -E|,|E_{n_0} -E|}
  \langle\Omega_{n,k}| \widehat{A} |\Omega_{n,k}\rangle,
\end{equation}
where $n_0$ is the index of an eigenvalue nearest to $E$.  Since
$M_{n_0}(E)$ is the number of terms in the last sum, the last expression
is nothing but the average of the expectation values of $\widehat{A}$ over
the energy eigenstates nearest to $E$. 

Therefore, whenever $E$ coincides with a point in the spectrum, the
limit $\varepsilon\to 0$ will give the microcanonical expectation value. 
On the other hand, if $E$ does not belong to the energy spectrum (in which
case the microcanonical ensemble is in principle ill-defined), then the
microcanonical result corresponding to the nearest eigenvalue is
obtained.  The only values of $E$ giving non-microcanonical limits are
those lying exactly in the middle between two eigenvalues, but even then the
result is an expectation value of a uniform distribution
over two energy eigenvalues. 

In the above, we tacitly assumed that it is possible to exchange the
order of the limit and the sum over $n$.  This operation is in
fact allowed if the sum in (\ref{e:tracesum}) {converges absolutely
for any $\varepsilon$}, which for positive $\widehat{A}$ is
tantamount to the assumption $\langle\widehat{A}\rangle_\varepsilon < \infty$.  
We will now prove this by showing that if (\ref{e:tracesum}) converges
absolutely for $\varepsilon_0>0$, then the series in (\ref{e:meanlim}) 
converges uniformly for $0<\varepsilon\leq\varepsilon_0$.    

Let $n_0$ be, as before, the index of one of the eigenvalues nearest to $E$.
Since $\exp(-{C\over \varepsilon^2})$ is an increasing function of
$\varepsilon$ for $C,\varepsilon > 0$, we immediately see that
\begin{equation}
\label{e:exineq}
{ 1 \over \sum\limits_{n',k'} \exp\!\left[ {
     -{(E_{n'} -E)^2 - (E_n -E)^2 \over 2 \varepsilon^2} }\right] } 
< \exp\!\left[ {-{(E_{n} -E)^2 - (E_{n_0} -E)^2 \over 2
      \varepsilon_0^2}}\right] 
\end{equation}   
for all $n$ such that $|E_{n} -E|> |E_{n_0} -E|$.  In fact,
this inequality extends also to the case $|E_{n} -E|= |E_{n_0} -E|$,
since the ratio on the left hand side is always less than one.
Therefore, the absolute value of 
each of the terms of the series in (\ref{e:meanlim}) is less than 
$\exp[{(E_{n_0} -E)^2 / (2 \varepsilon_0^2)}]
 \exp[{-(E_n -E)^2 / (2 \varepsilon_0^2)}] 
 |\langle\Omega_{n,k}| \widehat{A} |\Omega_{n,k}\rangle|$, which again form an
$\varepsilon$-independent sum that is convergent by assumption.  Thus the
series in (\ref{e:meanlim}) satisfies the Weierstrass $M$-test
and is uniformly convergent.

The uniform convergence of $\sqrt{2 \pi \varepsilon^2} \tr \tihep$
is similarly established by the Weierstrass $M$-test using the
monotonicity of $\exp[-(E_n -E)^2 / (2 \varepsilon^2)]$.

\section{Derivation of the lattice ensemble}

In this section we will derive a ``lattice approximation'' of the
Gaussian ensemble for quantum mechanical Hamiltonians of the type 
$\ham = {\widehat{p}^2\over 2 m} + V(\widehat{x})$,
where $V$ is a suitable potential function.  For example, if $V$ were a
(non-constant) polynomial bounded from below, then
$\ham$ would be essentially self-adjoint
and the operator $e^{-\beta \ham}$ would be trace-class with the
usual Feynman-Kac path-integral kernel.
Also, for any polynomial and for all bounded functions $A$, the canonical
expectation values would then be well-defined and could be computed by using the
formula\xcite{reedsimonII,simon:fint}
\begin{equation}\label{e:gbtrace}
\tr\!\!\left[A(\widehat{x})e^{-\beta\ham}\right] = \lim_{N\to\infty} 
 \int\!\! {d^N\!x\, d^N\!p \over (2\pi)^N}
  A(x_0) \exp\!\!\bigg[\ci \sum_{k=1}^N p_k (x_{k-1}-x_k) - 
  {\beta\over N} \sum_{k=1}^N\!\!\bigg({p_k^2\over 2 m} 
   + V(x_k)\bigg)\!\bigg],
\end{equation}
where $x_0=x_N$ and $\beta>0$.  

In the following we will need the analytic continuation of this result
to the region $\re \beta >0$.  If $V$ is a second-order polynomial, i.e.\
if we consider a harmonic oscillator, then this continuation can be
computed exactly and it is seen to be given by (\ref{e:gbtrace}).
There is no
apparent reason why the same continuation should not work for any
polynomial, but we have not yet been able to construct a
proof of this generality.\xcite{polybound} 
We will therefore simply assume that
$V$ is a potential such that (\ref{e:gbtrace}) holds for all $\re \beta >0$.

It will be useful to define a few auxilary quantities for the 
computation of the lattice expectation values.  Firstly, 
the relevant energy parameter of the lattice Gaussian distribution will turn
out to be $E/\vep$ and we will use a separate 
notation $W$ for this dimensionless quantity.  Secondly, we will need
the following three ``action-like'' quantities:
\begin{equation}\label{e:s0s1}
S_0(x;\vep) = {1\over N\vep} \sum_{k=1}^N {m\over 2}
  \left( {x_{k-1} - x_k \over 1/(N\vep) } \right)^2, \quad
S_1(x;\vep) = {1\over N\vep}\sum_{k=1}^N V(x_k)
\end{equation}
and, finally, $S(p,x; \vep)=  {1\over N\vep}\sum_{k=1}^N  
 {p_k^2 \over 2 m} + S_1(x;\vep)$. 
A comparision with (\ref{e:gbtrace}) reveals that $S$ is nothing 
but the usual finite-temperature lattice action for temperature $T=\vep$.

In order to keep track of all the mathematical details in the derivation 
of the lattice ensemble, we need to use a temporary regulator to provide
for absolute convergence of the integrals.  
In the following we have used $\exp(-\beta\ham)$ for this purpose---the
potentials have been chosen so that this is legitimate. 

Since the energy spectrum is bounded from below, we can use 
this lower bound to give a majorant in the M-test
and hence to justify the formula
\begin{equation}
\label{e:betastart}
 \tr(\tihep \widehat{A}) = \lim_{\beta\to 0^+}
\sum_{n,k} {1\over \sqrt{2 \pi \varepsilon^2} }
\exp\!\left[ -{1\over 2 \vep^2} \left( E_n - E \right)^2 -\beta E_n\right]
\langle\Omega_{n,k}| \widehat{A}|\Omega_{n,k}\rangle,
\end{equation}
for $\widehat{A}=A(\widehat{x})$ as before.
The Fourier transform of a Gaussian distribution will now give the
integral representation
\begin{equation}
\label{e:gaussint}
\int_{-\infty}^\infty {d\alpha\over 2\pi}
e^{-\half \varepsilon^2\alpha^2 + \ci \alpha E-\ci\alpha R} =
{1\over \sqrt{2 \pi \varepsilon^2} } e^{-{1\over 2 \vep^2}
{ (R-E)^2 } },
\end{equation}
which, by using the bound 
$\tr\!\!\left[|\widehat{A}|e^{-\beta\ham}\right] < \infty$
in the Lebesgue dominated
convergence theorem, will lead to the result
\begin{equation}
\label{e:lbegin}
\tr(\tihep \widehat{A}) = \lim_{\beta\to 0^+} \int_{-\infty}^\infty 
{d\alpha\over 2\pi}
e^{-\half \varepsilon^2\alpha^2 + \ci \alpha E}
\,\tr\!\!\left[e^{-(\beta+\ci\alpha)\ham} \widehat{A}\right].
\end{equation}

The trace in the integrand of (\ref{e:lbegin}) has now
a path-integral expression given by (\ref{e:gbtrace}).
Inserting the path-integral formula, referring once more to 
the dominated convergence theorem to move the ``continuum
limit'' over the $\alpha$-integration and applying the
Fubini theorem and eq.\ (\ref{e:gaussint}) will then yield
\begin{equation}
\label{e:betatrace}
\tr(\tihep \widehat{A}) = \lim_\beta \lim_N
\int\! {d^N\!x\, d^N\!p \over (2\pi)^N}
  { A(x_0) \over \sqrt{2 \pi \vep^2} }
   \exp\!\left[\ci \sum p_k (x_{k-1}-x_k) -\beta\vep S - \half (S-W)^2
   \right].
\end{equation}   
If we now use the equality
\begin{equation}
-\beta\vep S - {1\over 2} \left( S - W \right)^2
 = - {1\over 2} ( S - W + \beta\vep )^2 + 
   {1\over 2}\beta^2\vep^2 - \beta\vep W
\end{equation}
and take the limit of the last two $N$-independent terms,
the effect of using the regulator will reduce to a
shifting of the energy $E$ by
$-\beta\vep^2$.  But since the series in (\ref{e:betastart}) clearly
converges uniformly in $E$ and thus gives a continuous function in it,
the limit $\beta\to 0^+$ will amount to setting $\beta=0$ 
in (\ref{e:betatrace}).

Therefore, we have proved that
\begin{equation}
\label{e:ltrace}
\tr(\tihep \widehat{A}) = 
\lim_{N\to\infty} \int\! d^N\!x\, A(x_0) K_N(x; E,\vep),
\end{equation}
where the kernel is given by
\begin{equation}
K_N(x;E,\vep) = \int\!{d^N\!p \over (2\pi)^N} {e^{\ci \sum p_k (x_{k-1}-x_k)}
\over \sqrt{2 \pi \vep^2} }
 \exp\!\left[ -\half 
  \left( {p^2\over 2 m N \vep} + S_1(x;\vep) - W \right)^2 \right].
\end{equation}
Most of this integral can be evaluated by choosing
spherical coordinates with the $N$-axis pointing in the direction of
the vector $(x_{k-1}-x_k)$.  Let us denote by $y$ the rotated
$N$-coordinate and let us gather the remaining components into a
$N$$-$1-dimensional vector $q$. Then the kernel can be rewritten as
\begin{equation}
\label{e:kn2}
K_N = \int\!\!{d^{N-1}\!q \over (2\pi)^{N-1}} 
  \int_{-\infty}^\infty\!\!{dy\over 2\pi}
  {e^{\ci y \sqrt{\sum_k (x_{k-1}-x_k)^2}} \over \sqrt{2 \pi \vep^2} }
  \exp\!\left[ -\half \left( {y^2+ q^2\over 2 m N\vep} 
 	+ S_1 - W \right)^2 \right].
\end{equation}
Next we will choose polar coordinates in the $q$-space and rescale
both $q$ and $y$ by the factor $1/\sqrt{2 m N \vep}$.
Since the integrand is independent of the angles and the 
area of the unit sphere in $\mathbb{R}$$^n$ is given by
$2\pi^{n/2} \Gamma(n/2)^{-1}$, we will then arrive at the
formula
\begin{equation}
\label{e:knqy}
K_N = { 2 \left( mN\vep \right)^{N\over 2}
  \over (2\pi)^{N\over 2} \Gamma\!\left({N-1\over 2}\right) 
  \!\sqrt{2 \pi^2 \vep^2} }  
  \int_0^\infty\!\!\!dq\, q^{N-2} 
   \!\int_{-\infty}^\infty\!\!\!dy \, {\exp\!\left[ \ci 2 y \sqrt{S_0}
    -\half {\left( y^2 + q^2 + S_1 - W \right)^2}
   \right] },
\end{equation}
where we have made use of the definitions in (\ref{e:s0s1}).

It is possible to re-express this Fourier transform as a
one-dimensional real integral.  If we choose polar coordinates in
the $(y,q)$-plane, utilize the following 
integral representation for the Bessel functions of the first
kind,\xcite{besselint}
\begin{equation}
\int_0^\pi\!\! d\theta \sin^{N-2}\!\theta \,e^{\ci v \cos\theta} =
 \Gamma\!\left({N-1\over 2}\right) \sqrt{\pi} 
   ( 2 / v )^{N-2\over 2} J_{N-2\over 2}(v),
\end{equation}
and, finally, denote $r=y^2+q^2$ and $\nu={N-2\over 2}$, we have
\begin{equation}
\label{e:Kint}
K_N = \left( {m N\vep \over 2 \pi}\right)^{N\over 2} 
  {1 \over \sqrt{2 \pi \vep^2} }
   \int_0^\infty\! dr \left( {r \over S_0}\right)^{\nu\over 2} 
  \! J_\nu\!\left(2\sqrt{S_0 r}\right) \exp\!\left[-\half \left(r + S_1- 
     W \right)^2 \right].
\end{equation}

The integral representation (\ref{e:Kint}) can now be used to derive a
series expansion for the kernel.  Since 
\begin{equation}
 \left( {r \over S_0}\right)^{\nu\over 2} \! 
 J_\nu\!\left(2\sqrt{S_0 r}\right) =
 \sum_{n=0}^\infty {(-S_0)^n r^{\nu+n} \over n!\, \Gamma(\nu+n+1)},
\end{equation}
the expansion of the exponential $\exp[-r(S_1-W)]$ and a subsequent
term by term integration will give
\begin{equation}
K_N = \left( {m N\vep \over 2 \pi}\right)^{N\over 2}
  { \exp\!\left[ -\half \left(S_1-W \right)^2 \right] \over
   \sqrt{2 \pi \vep^2} }
  \sum_{n=0}^\infty \sum_{k=0}^n
     { (-S_0)^k \over k!} { \left(W-S_1\right)^{n-k}  \over (n-k)!}  
      {2^{\nu-1+n\over 2} \Gamma\!\left({\nu+1+n\over 2}\right) \over
        \Gamma(\nu+1+k)},
\end{equation}
which, however, has no immediate applications for evaluation of the kernel,
because the convergence of
the sum will be too slow for large values of the parameters. 

So far, the best way to evaluate the kernel $K_N$ for large enough range
of parameter values has been a numerical integration of eq.\
(\ref{e:Kint}). This is mainly due to the well-developed methods of
computation of the values of the Bessel functions.
Similar methods---recursion relations, integral representations,
etc.---will likely offer more efficient ways of evaluating the kernel 
as well. However, since the
numerical integration was sufficiently fast for our purposes, we
do not pursue this point further here.

\section{Numerical evaluation of the lattice expectation 
  values}\label{sec:num}

The purpose of this section is to enable a Monte Carlo integration of
the lattice expectation values in (\ref{e:ltrace}).  From the experience in
the lattice quantum field simulations, this should be possible provided
that the following condition is met: the Monte Carlo is used for a
generation of a probability distribution $\exp(-S[x])$, where $S[x]$ is
a {\em local}\/ action.  The locality, which means that the action is a
sum over a local ``action density'', makes it possible to use local updating
algorithms and in this way to achieve a fast enough generation of the required
probability distribution.  In the following we will derive an 
exponential bound for the kernel, but with an ``action'', which is
not local.  However, as explained at the end of this section, a certain
linear approximation for the exponent will be a local quantity, using
which it is still possible to benefit from the local
updating algorithms. 

Letting $r=q^2$ in (\ref{e:knqy}) will give the starting point for the
approximations:
\begin{equation}
\label{e:knstart}
K_N = { \left( mN\vep \right)^{N\over 2}
  \over (2\pi)^{N\over 2} \Gamma\!\left({N-1\over 2}\right)
  \sqrt{2 \pi^2 \vep^2} }
  \int_0^\infty\!\!dr\, r^{N-3\over 2}
   \!\int_{-\infty}^\infty\!\!dy \, {\exp\!\left[ \ci 2 y\sqrt{S_0}
    -\half {\left( y^2 +r + S_1 - W \right)^2}
   \right] }.
\end{equation}
Next we need to rely on the lemmas proved in the Appendix.  First on
Lemma \ref{th:cmpli}, which states 
that for all real $\lambda$, the $y$ in the integrand of (\ref{e:knstart})
can be replaced by $y$$+$$\ci \lambda$, i.e.\ that the double integral 
is always equal to
\begin{equation}
\label{e:ryint}
  \int_0^\infty\!\!dr\, r^{N-3\over 2}
   \!\int_{-\infty}^\infty\!\!dy \, {\exp\!\left[ \ci 2 y\sqrt{S_0}
   -2\lambda\sqrt{S_0} -\half \left( y^2 -\lambda^2 
      +r + S_1 - W +\ci 2 \lambda y \right)^2 \right] }.
\end{equation}
The absolute value of this quantity is less than
\begin{equation}
  \int_{-\infty}^\infty\!\!dy \,
    \exp\!\left[-2\lambda \sqrt{S_0} + 2 \lambda^2 y^2 \right]
  \! \int_0^\infty\!\!dr\, r^{N-3\over 2}
   \, {\exp\!\left[ -\half \left( r + y^2 -\lambda^2 
      + S_1 - W \right)^2 \right] }.
\end{equation}

Now we are ready to apply Lemma \ref{th:rnubound} 
to get an upper bound for the second integral,
\begin{equation}\label{e:ryint2}
  \int_{-\infty}^\infty\!\!dy \, 
    \exp\!\left[ -2\lambda \sqrt{S_0} + 2 \lambda^2 y^2 \right]
     { \sqrt{\pi} \Gamma\!\left({N-1\over 2}\right)
        \over 2^{N-1 \over 4} \Gamma\!\left({N+1\over 4}\right) }
      \exp\!\left[ -b_N \left( y^2 - \lambda^2 + S_1 - W \right) \right],
\end{equation}
with $b_N = \sqrt{2} \Gamma\!\left({N+1\over 4}\right) /
\Gamma\!\left({N-1\over 4}\right)$ (for large values of $N$, $b_N$ can
be approximated by $\sqrt{N-2\over 2}=\sqrt{\nu}$). Since
\begin{equation}
\int_{-\infty}^\infty\!\!dy\, \exp\!\left[ - (b_N - 2 \lambda^2) y^2\right]
 = \sqrt{\pi \over b_N - 2 \lambda^2} \le \sqrt{\pi},
\end{equation}         
whenever $2 \lambda^2 \le b_N-1$, we can infer from (\ref{e:ryint2})
that for these values of $\lambda$
\begin{equation}
|K_N| \le  \left( {mN\vep \over 2\pi} \right)^{N\over 2}
  {1 \over \sqrt{2 \pi \vep^2} }
  { \sqrt{\pi} \over 2^{N-1 \over 4} \Gamma\!\left({N+1\over 4}\right) }
   \exp\!\left[ -2\lambda \sqrt{S_0} + b_N \lambda^2 -b_N
     (S_1 - W) \right].
\end{equation}
As a function of $\lambda$, the right hand side has only one minimum
which lies at $\lambda = \sqrt{S_0} / b_N$.  If $S_0 \le b_N^2 (b_N-1)/2$,
we can use this value of $\lambda$ to give the
upper bound, and the exponential factor will become $\exp\!\left[ -
S_0/b_N - b_N (S_1 - W) \right]$.

Let us now use notation $M_N$ for the critical value of $S_0$---for
large $N$, $M_N$ is approximately equal to $\sqrt{\nu^3}/2$.  
If $S_0 > M_N$, the best allowed value is then given by $\lambda =
\sqrt{ b_N-1 \over 2} = \sqrt{M_N}/b_N$, which yields
the exponential factor
$\exp\!\left[ -(2 \sqrt{M_N S_0} - M_N)/b_N - b_N (S_1 - W) \right]$.  

To summarize, we have now proved that
\begin{multline}\label{e:ubound}
|K_N|  \le \left( {mN\vep \over 2\pi} \right)^{N\over 2} 
  {1 \over \sqrt{2 \pi \vep^2} }   { \sqrt{\pi} \over 2^{N-1 \over 4}
     \Gamma\!\left({N+1\over 4}\right) }
     \exp\!\left[ W b_N - \left({S_0\over b_N} + b_N S_1 \right) \right] \\
    \times \exp\!\left[ \Theta\!\left(S_0-M_N\right) 
         {\left(\sqrt{S_0}-\sqrt{M_N}\right)^2 \over b_N} \right],
\end{multline}
where $\Theta$ denotes the step function.  As a curiosity, it should be
noted that the term ${S_0/ b_N} + b_N S_1$ corresponds to a
certain ``renormalized'' finite temperature lattice action---renormalized
in the sense that the temperature of the action depends on
the size of the lattice: $T(N)=\vep / b_N$.  

We propose that the lattice ensemble expectation values 
could be evaluated by using a Monte Carlo sampling to generate a 
distribution $\exp(-S^{(N)}(x;E,\vep))$ and thus allowing the evaluation of 
the ratios\xcite{footn1}
%\footnote{Note that all $x$-independent scaling factors
%coming from the normalization of the probability distribution
%$\exp(-S^{(N)})$ vanish in the ratio.}
\begin{equation}
{ \int\! d^N\!x \, A(x_0) \tilde{K}_N(x;E,\vep) \exp(-S^{(N)}(x;E,\vep))
   \over \int\! d^N\!x\,\tilde{K}_N(x;E,\vep) \exp(-S^{(N)}(x;E,\vep))
},
\end{equation}
where $\tilde{K}_N$ is obtained from the 
scaling of the kernel by the inverse of the right hand side of 
(\ref{e:ubound}) and the ``lattice action'' $S^{(N)}$ is given by the
formula
\begin{equation}
S^{(N)} = {S_0 - \Theta\!\left(S_0-M_N\right)
  \left(\sqrt{S_0}-\sqrt{M_N}\right)^2 \over b_N} + 
  b_N\left(S_1 - {E/\vep}\right).
\end{equation}
For local updating algorithms, the value of the action could be computed
exactly at the beginning of each sweep and the linear approximation
\begin{equation}
\Delta S^{(N)} = %\sqrt{M_N \over M_N + (S_0-M_N) \Theta(S_0-M_N)}
{c\over b_N} \Delta S_0 + b_N \Delta S_1  
\end{equation}
could be used during the sweep.  In the linear approximation,
$c=1$, if $S_0 \le M_N$ at the beginning of the sweep, and
$c=\sqrt{M_N / S_0}$ otherwise.

To check that the approximations used in the derivation of
the upper bound are reasonable, 
we have shown a plot of values of the normalized kernel
$\tilde{K}_N$ with $N=22$ and $W=7$ in Fig.\ 1\@.
Two general features are evident from the figure:
firstly, there is a ``gaussian wall'' around the line $S_1=W$, which means
that the importance sampling should extend at least that far in the
$S_1$-direction.
Secondly, although the scaling can be seen to be quite optimal for the region 
$S_0 \le M_N$, the scaled values drop rapidly after that point.  
The oscillations of the kernel, however, also begin at
about the same values of $S_0$ and an oversampling 
of the oscillatory region is in fact
only desirable---therefore, the use of this particular
scaling function is qualitatively justified.

\section{Conclusions}
We have defined a Gaussian quantum statistical
ensemble for the approximation of microcanonical expectation
values of any system which is amenable for the usual Gibbs ensemble
treatment. We have also derived a method for the computation of
these expectation values by a Monte Carlo integration
on a time-lattice, provided that the Hamiltonian allows an
analytic lattice regularization.  The generalization of these results
for quantum field theories and tests for the numerical efficiency of the
lattice approximation are currently under investigation.

\acknowledgements

I would like to thank Masud Chaichian, Antti Kupiainen, 
Claus Montonen and Esko Suhonen 
for their scientific guidance and support
and Sami Virtanen for several useful discussions.

\appendix
%\section{Two Lemmas}
\section{}

In this Appendix we shall prove the two lemmas that where used in
Section \ref{sec:num} to derive the exponential bound for the integral
kernel. 
\begin{lemma}
\label{th:cmpli}
Let $f$ be defined by the
formula $f(y) = \exp\!\left[\ci b y  -\half(y^2 + C)^2\right]$, where
$C$ and $b$ are complex parameters.
The integrals of $f$ over lines parallel to the real axis are all equal:
\[
 \int_{-\infty}^\infty\!\!dy\, f(y+\ci\lambda)
 = \int_{-\infty}^\infty\!\!dy\, f(y)\text{, for all }\lambda\in \mathbb{R}.
\]
\end{lemma}
\begin{trivlist}\item[]{\bf Proof.}
Consider the integral of $f$ over a countour following
the perimeter of a square with corners at points
$\pm R$, $\pm R+ \ci \lambda$.  Since $f$ is a composition of two entire
functions, it is entire and the integral over this countour is therefore 
zero by the Cauchy theorem.  Writing this integral in a parameter form
gives
\[
\int_{-R}^R\!\!dy\, f(y) = \int_{-R}^R\!\!dy\,
 f(y+\ci\lambda) - \ci \lambda \int_0^1\!\! dt 
 (f(R+\ci\lambda t) - f(-R+\ci\lambda t) ).
\]
But since for all $0\le t \le 1$,
\[
|f(\pm R+\ci\lambda t)| \le \exp\!\left[ |bR|+|b\lambda| - \half
  (R^2 - \lambda^2 - |C|)^2 + \half (2|\lambda|+|C|)^2 \right],
\]
and the upper bound goes to zero when $R$ goes to infinity, the
integral over $t$ vanishes when we take the limit $R\to \infty$ 
and the lemma follows. \mbox{$\Box$}
\end{trivlist}
\begin{lemma}
\label{th:rnubound}
For all $\tau > 0$ and $C\in \mathbb{R}$
\[
\int_0^\infty\!\!dr \, r^\tau e^{-\half (r+C)^2} 
 \le a_\tau e^{-b_\tau C},
\]
where $a_\tau$ and $b_\tau$ are positive constants given by the
formulae 
\[
 a_\tau = { \sqrt{\pi} \Gamma(\tau+1) 
  \over 2^{\tau+1\over 2} \Gamma\!\left( {\tau+2 \over 2} \right)}, \qquad
 b_\tau = { \sqrt{2} \Gamma\!\left( {\tau+2 \over 2} \right)
  \over \Gamma\!\left( {\tau+1 \over 2} \right) }.
\]
\end{lemma}
\begin{trivlist}\item[]{\bf Proof.}
Let us use the notation $I_\tau(C)$ for the integral
$\int_0^\infty\!\!dr \, r^\tau e^{-\half (r+C)^2}$.  The integral
converges for $\tau > -1$, when $I_\tau$ defines a function 
from the real numbers to $(0,\infty)$.  We claim that the following
three properties are satisfied for all $C$, if $\tau > 0$:
\newcounter{alp}
\renewcommand{\thealp}{\alph{alp}}
\begin{list}{(\thealp)}{\usecounter{alp}}
\item\label{c:recurs} $I_{\tau+1}(C) = \tau I_{\tau-1}(C)-C I_\tau(C)$,
\item\label{c:deriv} $I_\tau$ is differentiable at $C$ and
  $I'_\tau(C) = -I_{\tau+1}(C) - C I_\tau(C) = -\tau I_{\tau-1}(C)$,
\item\label{c:Iratio} ${I_{\tau+1}(C) \over I_\tau(C)} >
    {I_\tau(C) \over I_{\tau-1}(C)}$.
\end{list}
The lemma is then a fairly straightforward consequence of these results.

Fix $\tau > 0$ and define $g(C) = \ln I_\tau(C)$.  Applying
(\ref{c:deriv}) twice shows that the first and the second derivatives of $g$
exist and that they are $g'= -\tau {I_{\tau-1}/ I_\tau}$
and $g'' = -\tau ({I_{\tau+1}\over I_\tau} {I_{\tau-1}\over I_\tau} -1)$.
But from (\ref{c:Iratio}) we now deduce that $g'' < 0$ and therefore
$g$ is a concave differentiable function.  This means that every tangent
line to $g$ is an upper bound for the function: for every
fixed $D\in\mathbb{R}$, $g(C) \le (C-D) g'(D) + g(D)$.  Exponentiation
of both sides of this inequality gives now
\[
I_\tau(C) \le I_\tau(D) 
 \exp\!\left[-\tau {I_{\tau-1}(D) \over I_\tau(D)} (C-D)\right],
\]
which is of the desired form if we just define
$b_\tau = \tau I_{\tau-1}(D) / I_\tau(D)$ and 
$a_\tau = I_\tau(D) \exp[b_\tau D]$.

It now remains to show that the explicit formulas given in the Lemma
for $a_\tau$ and $b_\tau$ are valid for some $D$.  Choose $D=0$, when a
change of variables from $r$ to $\half r^2$ shows that
$I_\tau(0)= 2^{\tau-1\over 2} \Gamma\!\left( {\tau+1\over 2}\right)$.
On the other hand, 
an application of (\ref{c:recurs}) reveals that for $D=0$,
$b_\tau = I_{\tau+1}(0)/ I_\tau(0)$, which then simplifies into 
the desired formula.  Since now $a_\tau = I_\tau(0)$, 
the form for $a_\tau$ is a consequence of the
Gamma-function ``doubling formula'': 
$\Gamma\!\left( {z\over 2 }\right) \Gamma\!\left( {z+1\over 2}\right)
= 2^{1-z} \Gamma\!\left(\half\right)\Gamma\!\left( z \right)$.

Therefore, it is necessary only to show that the propositions (a)--(c)
hold.  (\ref{c:recurs}) is obtained by a simple partial integration, since
$r\exp[-\half(r+C)^2] = - {d\over dr} \exp[-\half(r+C)^2] - C
\exp[-\half(r+C)^2]$ and the boundary terms vanish for $\tau > 0$.

Consider next the finite difference term, $I_\tau(C+h) - I_\tau(C)$.  
This is equal to $\int_0^\infty\!\!dr \, r^\tau e^{-\half (r+C)^2}
 \{\exp[-\half h^2-h (r+C)] -1\}$, where the term in the braces is
bounded by $\exp[ |h| r + |h C|]+1$. Lebesgue dominated convergence
theorem now shows that it is possible to differentiate under the integral
sign, which will prove the differentiability and the first equality in
(\ref{c:deriv}).  The second equality follows then by an application of
(\ref{c:recurs}). 

To prove (\ref{c:Iratio}), it is enough to show that 
$I_{\tau+1}(C) I_{\tau-1}(C) - I^2_\tau(C) > 0$, for all $C$ and $\tau
> 0$.  Expressing the products as two-dimensional integrals will yield
\[
I_{\tau+1}(C) I_{\tau-1}(C) - I^2_\tau(C) =
\int_0^\infty\!\!dx \int_0^\infty\!\! dy\, x^\tau y^{\tau -1} (x-y)
\exp\!\left[-\half\left( (x+C)^2+(y+C)^2 \right)\right].
\]
In the change of variables $(x,y)\to (t,u) = (x+y, {x-y\over x+y})$,
%$t=x+y$ and $u={x-y\over x+y}$
the Jacobian equals $\half t$ and the integration region changes into 
$(0,\infty)\times (-1,1)$.  This way the integral becomes
\[
\int_0^\infty\!\!dt \int_{-1}^1\!\! du\, 2^{-2 \tau} t^{2\tau +1} (u + u^2)
(1-u^2)^{\tau-1} \exp\!\left[-{1\over 4} t^2 (1+u^2) - C t -C^2 \right].
\]
As the region of integration over $u$ is symmetric, the 
antisymmetric part vanishes, and we arrive at the result
\[
I_{\tau+1}(C) I_{\tau-1}(C) - I^2_\tau(C) =
\int_0^\infty\!\!dt \int_{-1}^1\!\! du\, 2^{-2 \tau} t^{2\tau +1} 
u^2 (1-u^2)^{\tau-1} \exp\!\left[-{1\over 4} t^2 (1+u^2) - C t -C^2
\right],
\]
which is always strictly positive. \mbox{$\Box$}
\end{trivlist}


\begin{references}
\bibitem{polybound}
It is fairly easy to prove that the left hand side of
(\protect\ref{e:gbtrace}) is an analytic function in the region 
$\re\beta > 0$, if $V$ is a polynomial bounded from below.  
If we also assume that $A$ is a polynomial,
the analyticity of the right hand side will follow from an application
of the Cauchy theorem, but only
in the sector $|{\rm Arg}\ \beta|< {\pi \over n+2}$, where $n$ is the
degree of $V$.
\bibitem{footn1}
Note that all $x$-independent scaling factors,
coming from the normalization of the probability distribution
$\exp(-S^{(N)})$, vanish in the ratio.
\end{references}
\end{document}